\documentclass[twocolumn,showpacs,preprintnumbers,amsmath,amssymb]{revtex4} 
\usepackage{graphicx}% Include figure files 
\usepackage{dcolumn}% Align table columns on decimal point 
\usepackage{bm}% bold math 
 
\begin{document} 
 
\preprint{APS/123-QED} 
 
\title{Unconventional Superconductivity in the Novel Layered Superconductor Fe(Te-Se)
\\ Investigated by $^{125}$Te NMR on the Single Crystal}% Force line breaks with \\ 
 
\author{Chishiro Michioka$^1$} 
\email{michioka@kuchem.kyoto-u.ac.jp} 
\author{Hiroto Ohta$^1$}
\author{Mami Matsui$^1$} 
\author{Jinhu Yang$^1$}
\author{Kazuyoshi Yoshimura$^1$} 
\email{kyhv@kuchem.kyoto-u.ac.jp} 
\author{Minghu Fang$^2$} 
\affiliation{$^1$Department of Chemistry, Graduate School of Science, Kyoto 
University, Kyoto 606-8502, Japan} 
\affiliation{$^2$Department of Physics, Zhejiang University, Hangzhou 310027, 
China}

\date{\today}% It is always \today, today, 
 
\begin{abstract}
We have performed $^{125}$Te NMR on a single crystal of the novel layered 
superconductor Fe$_{1.04}$Te$_{0.67}$Se$_{0.33}$ for the first time. The spin parts 
of the Knight shifts for both $H // a$, $H // c$ are suppressed in the superconducting state, 
indicating the spin singlet superconductivity. At superconducting state, 
$1/T_{1}$ shows the power-law behavior ($\sim T^{3}$) without any coherent peaks. Observations of the residual 
density of state and the $T^{3}$-law indicate the presence of the line node in the 
superconducting gap. In the normal state, $1/T_{1}T$ which probes the \textbf{q}-summation of spin fluctuations 
enhances at low temperatures. Our results suggest that the superconductivity in 
Fe$_{1+\delta}$Te$_{1-x}$Se$_{x}$ is stabilized by the growth of the antiferromagnetic 
spin fluctuations as well as in FeSe. 
\end{abstract}
 
\maketitle 
 
Since recent discoveries of iron based superconductors with the superconducting 
transition temperature ($T_{\textrm{c}}$) being 55 K in the highest cases \cite{Kamihara, Ren, Chen, 
Kito},  many studies have been done to search new materials with high 
$T_{\textrm{c}}$ and to clarify the mechanism of superconductivity. In such 
investigations, the superconductivity was discovered with $T_{\textrm{c}}$ = 8 K in the $\alpha$-FeSe 
system  \cite{Hsu}. The $\alpha$-FeSe has a simple layered structure, in 
which the tetragonal FeSe layers stack continuously along the $c$-axis without insertion of another layers. 
However, the presence of Se defects are discussed in the report of the discovery of the 
superconductivity \cite{Hsu}, Williams $et$ $al.$ indicated that the stoichiometric 
FeSe is the most preferable for the superconductivity \cite{Williams}, since the excess 
irons should suppress the superconductivity \cite{McQueen}. In the next stage, 
superconductivity in Se-substituted (maximum $T_{\textrm{c}}$ = 14 K\cite{Fang, 
Yeh}) and S-substituted (maximum $T_{\textrm{c}}$ = 10 K\cite{Mizuguchi}) systems at 
the Te sites in FeTe which is isostructural with the superconducting FeSe 
were discovered. Fang $et$ $al.$ presented that the end member $\alpha$-FeTe shows 
a magnetic phase transition at 65 K and the superconductivity of 
FeTe$_{1-x}$Se$_x$ occurs when the magnetic phase transition is suppressed by the 
increase of $x$ \cite{Fang}. Because these Fe chalcogenide systems FeSe and 
FeTe$_{1-x}$Se$_x$ layers have the simplest structure of stacked quasi-two 
dimensionally, we can regard them as important key compounds to clarify 
the intrinsic properties of Fe-based superconductors.

In all Fe-based superconductors, the superconductivity occurs in the vicinity of the 
magnetic phase. Therefore, the magnetic fluctuations are thought to play an 
important role in the Fe-based superconductors. NMR is a powerful probe to investigate
low energy excitations of fluctuated spins. Imai $et$ $al.$ presented that in almost 
stoichiometric FeSe, the relaxation rate divided by temperature ($1/T_{1}T$) which is 
proportional to the \textbf{q}-summation of the imaginary part of the dynamical spin susceptibility 
$\textrm{Im}\chi(\textbf{q})$ is enhanced at low temperatures \cite{Imai}. In a 
slightly Fe-rich FeSe$_{0.92}$ sample, the short $T_{1}$ component $1/T_{1S}T$ also 
shows the enhancement with decreasing temperature together with almost constant $1/T_{1L}T$ of the 
long $T_{1}$ component \cite{Kotegawa, Masaki}. From a point of view of electron 
spin fluctuations, there would be two-types in Fe-based superconductors: one shows the 
enhancement of $1/T_{1}T$ at low temperatures and the other does not show such the 
enhancement. The former group includes the 122-system (e.g. 
Ba$_{1-x}$K$_x$Fe$_2$As$_2$)\cite{Baek, Fukazawa, Ning1, Ning2} and the 
11-system (e.g. FeSe)\cite{Imai, Masaki}, and the latter the 1111-system (e.g. 
LaFeAsO$_{1-x}$F$_x$)\cite{Nakai1, Nakai2, Mukuda}. Even if there are some 
differences in each compound, it should be quite natural to consider that all the Fe based 
superconductors have the same superconducting mechanism. It is therefore very important to 
investigate the spin fluctuations and the antiferromagnetic quantum criticality of the Fe 
based superconductors. Our purpose is to clarify the role of spin fluctuations from the 
antiferromagnetic FeTe to superconducting FeTe$_x$Se$_{1-x}$ systematically. In this letter 
we present $^{125}$Te NMR studies in the single crystal 
Fe$_{1.04}$Te$_{0.67}$Se$_{0.33}$ for the first time.

The Te site in antiferromagnet FeTe can be substituted by Se in all the composition range 
\cite{Fang,Yeh}, and excess irons are doped in the other Fe site as nonstoichiometric 
Fe$_{1+\delta}$Te$_{x}$Se$_{1-x}$ \cite{Fruchart, Bao}. Since there is a variety of nonstoichiometry, 
characterizations of samples are so important. 

The single crystal of Fe$_{1.04}$Te$_{0.67}$Se$_{0.33}$ was prepared by the self-flux method 
in an evacuated quartz tube. The sample used for NMR measurements is the same as the previous 
report \cite{Yang}. The sample was characterized by X-ray diffraction, and the 
compositions were determined by energy dispersive spectroscopy. The crystal axes 
were also decided by Laue photographs. All NMR measurements were carried out by using 
spin-echo method with a standard phase coherent-type NMR pulsed spectrometer applying  
the magnetic field $H$ parallel to the $a$ or $c$-axis. NMR spectra were measured with sweeping magnetic 
field in a constant NMR frequency. The nuclear spin-lattice relaxation time ($T_{1}$) was measured 
by inversion recovery method. $^{125}$Te nucleus with the nuclear spin $I$ = 1/2 has the nuclear gyromagnetic 
ratio $\gamma$ = 13.454 MHz/T. The magnetic field was calibrated by using reference signals of $^{2}$D in 
D$_{2}$O and $^{125}$Te in TeCl$_{4}aq$.

In the Fe$_{1+\delta}$Te$_{x}$Se$_{1-x}$ system, excess irons are thought to 
suppress the superconductivity, e.g.,  the upper critical field ($H_{\textrm{c2}}$), 
and this situation is similar to the case of Fe$_{1+\delta}$Se \cite{Yang, McQueen}. 
Fe$_{1.04}$Te$_{0.67}$Se$_{0.33}$ shows large superconducting volume fraction 
while Fe$_{1.12}$Te$_{0.72}$Se$_{0.28}$ shows small fraction\cite{Yang}. From 
the results of the electric resistivity measurements, both samples show the onset 
$T_{\textrm{c}}$ = 15 K, while the critical temperature estimated from zero resistivity in 
Fe$_{1.04}$Te$_{0.67}$Se$_{0.33}$ is higher than that in 
Fe$_{1.12}$Te$_{0.72}$Se$_{0.28}$ \cite{Yang}. 

%fig 1
\begin{figure}[tb]
\begin{center}
\includegraphics[width=7cm]{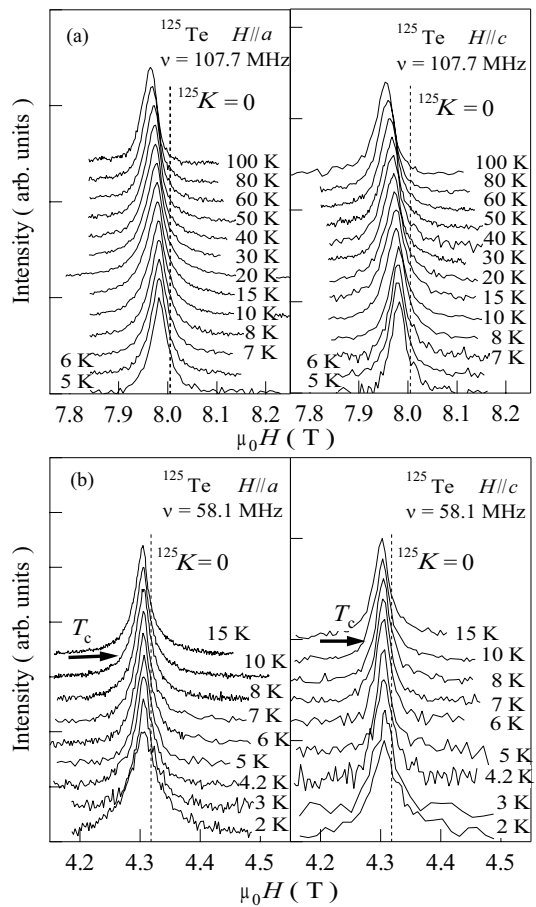}
\end{center}
\caption{
Field-swept $^{125}$Te NMR spectra in Fe$_{1.04}$Te$_{0.67}$Se$_{0.33}$ at constant frequencies (a) $\nu = 107.7$ 
MHz and (b) $\nu = 58.1$ MHz with magnetic fields $H$ parallel to the $a$ and  $c$-axes.
}
\label{fig:spe}
\end{figure}

Figure \ref{fig:spe} shows the field-swept $^{125}$Te NMR spectra of Fe$_{1.04}$Te$_{0.67}$Se$_{0.33}$. In Fig. \ref{fig:spe}-(a), we show the spectra in the normal state in a constant frequency $\nu$ = 107.7 MHz (at about 8 T). In both conditions of $H //a$ and $H // c$, the central frequencies of the spectra increase with decreasing temperature. The linewidth is about three times broader than that in Fe$_{1.01}$Se \cite{Imai}, indicating distribution of the local electronic state around Te caused by the Se-substitution as well as the presence of excess irons. In our preliminary investigations, the linewidth of the NMR spectrum decreases with the decrease of the amount of excess irons which produce nearly-localized Curie-Weiss-like temperature dependence in the magnetic susceptibility. The spectra around and below $T_{\mathrm{c}}$ are shown in Fig. \ref{fig:spe}-(b). As well as the normal state, the central frequencies for both $H //a$ and $H // c$ increase slightly with the decrease of temperature. Below $T_{\mathrm{c}}$ the spin-echo intensity of the NMR spectrum is markedly weakened. The linewidth of the spectrum increases with decreasing temperature accompanied by the decrease of the London 
penetration depth $\lambda$. 

%fig 2
\begin{figure}[tb]
\begin{center}
\includegraphics[width=7cm]{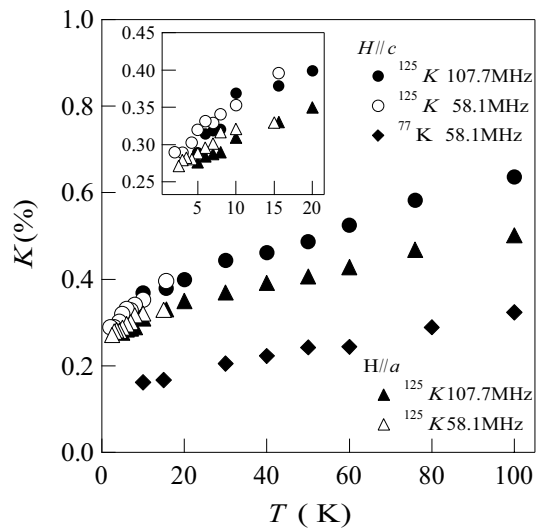}
\end{center}
\caption{
Temperature dependence of Knight shifts, $^{125}K$ and $^{77}K$ in 
Fe$_{1.04}$Te$_{0.67}$Se$_{0.33}$. 
}
\label{fig:Kall}
\end{figure}

 We show the temperature dependence of the Knight shift $^{125}K$ in 
Fe$_{1.04}$Te$_{0.67}$Se$_{0.33}$ in Fig. \ref{fig:Kall} with the Knight shift of  
$^{77}$Se nucleus $^{77}K$ for comparison. In the normal state, both $^{125}K$ 
and $^{77}K$ decrease with decreasing temperature, similar to the case with
Fe$_{1.01}$Se \cite{Imai}. The uniform spin susceptibility ($\chi$) in
Fe$_{1+\delta}$Te$_{1-x}$Se$_{x}$ frequently shows Curie Weiss-like behavior at 
low temperatures, and the Curie constant depends on the amount of the excess irons, 
$\delta$ \cite{Yang}. In the case of the sample which contains very few excess irons, 
$\chi$ does not show such the Curie Weiss-like behavior, and decreases with 
decreasing temperature with scaled to $K$. This behavior of the intrinsic spin susceptibility 
is characteristic of the case with itinerant antiferromagnets, i.e., $\chi_{0}$ is suppressed by 
the growth of $\chi(\mathbf{q})$ \cite{Imai, Johnston, Takigawa}. 

%fig 3
\begin{figure}[tb]
\begin{center}
\includegraphics[width=7cm]{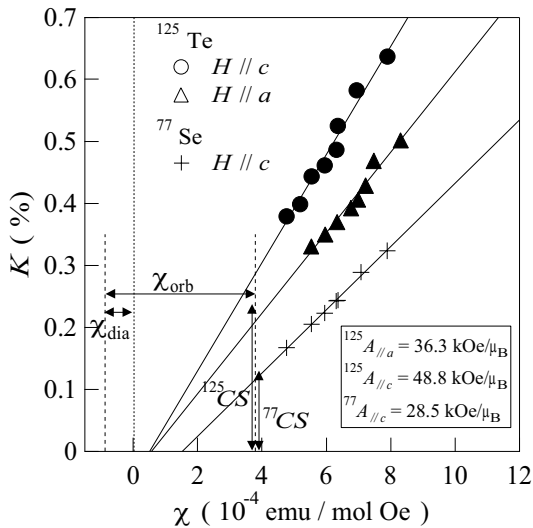}
\end{center}
\caption{ Knight shifts $^{77}K$ and $^{125}K$ plotted against the uniform 
susceptibility $\chi$ in Fe$_{1.04}$Te$_{0.67}$Se$_{0.33}$. 
$\chi\mathrm{_{dia}}$, $\chi\mathrm{_{orb}}$, $^{77}CS$ and 
$^{125}CS$ denote diamagnetic and orbital contributed $\chi$ and chemical shifts of 
$^{77}$Se and $^{125}$Se nuclei, respectively.
}
\label{fig:Kchi}
\end{figure}

Figure \ref{fig:Kchi} shows Knight shifts plotted against the uniform 
susceptibility (so called $K-\chi$ plot) of $^{77}$Se and $^{125}$Te nuclei in 
Fe$_{1.04}$Se$_{0.33}$Te$_{0.67}$ with temperature as an implicit parameter. As 
discussed above, when the amount of excess irons is small, $\chi$ does not show the 
Curie-Weiss like enhancement and decreases with decreasing temperature similar to 
$K$. The $K-\chi$ plots show good linearity as seen in Fig. \ref{fig:Kchi}. From the slopes, we estimated 
hyperfine coupling constants as $^{125}A_{//a}$ = 36.3 kOe/$\mu\mathrm{_{B}}$, 
$^{125}A_{//c}$ = 48.8 kOe/$\mu\mathrm{_{B}}$ and $^{77}A_{//c}$ = 28.5 
kOe/$\mu\mathrm{_{B}}$. The temperature dependent $K$ and $\chi$ originate in the spin part. 
Next, we consider the temperature-independent terms of $K$ and $\chi$.
In general, the chemical shift of $^{125}$Te nucleus is large, sometimes up to 0.3 $\%$, 
and is by a factor of $\sim$ 2 larger than that of $^{77}$Se nucleus which is located in the same chemical 
surroundings \cite{chemicalshift}. From the ordinary ratio of chemical shifts in $^{125}$Te and $^{77}$Se in the same site, 
we can roughly estimate the chemical shifts in Fe$_{1.04}$Se$_{0.33}$Te$_{0.67}$  as 
$^{125}CS \sim 0.25 \%$ and $^{77}CS \sim 0.12 \%$, and then $\chi\mathrm{_{orb}}$ can be estimated as 
$\sim 4.5 \times 10^{-4}$ emu/mol as shown in Fig. \ref{fig:Kchi}.

 In the superconducting state, both $^{125}K_{a}$ and $^{125}K_{c}$ 
decrease with decreasing temperature. From the roughly estimated chemical shifts in Fe$_{1.04}$Se$_{0.33}$Te$_{0.67}$, 
which gives us proper estimatimation of $\chi\mathrm{_{orb}}$, the spin part of the Knight shift $K\mathrm{_{spin}}$ seems to be 
completely suppressed in the ground state for both magnetic field directions, indicating 
the spin-singlet superconductivity. Since the spectrum shape changes at low 
temperatures, we need detailed studies at more low field. 

%fig 4
\begin{figure}[tb]
\begin{center}
\includegraphics[width=7cm]{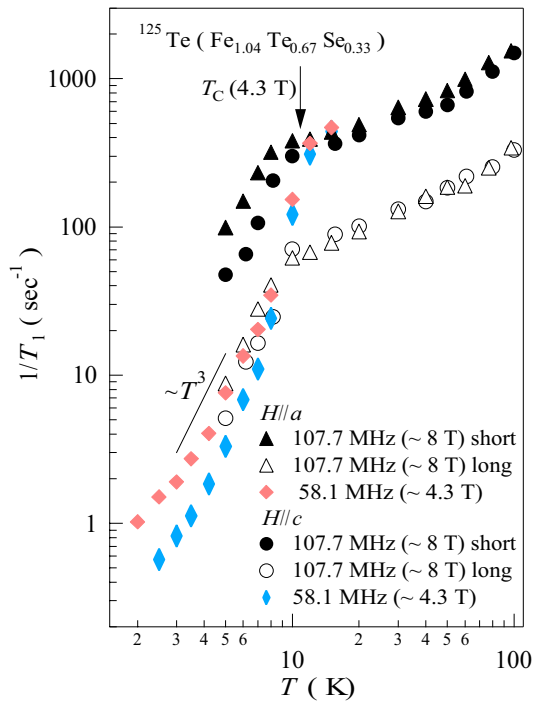}
\end{center}
\caption{
(a)Temperature dependence of the nuclear spin-lattice relaxation rate ($1/T_{1}$) of $^{125}$Te in 
Fe$_{1.04}$Te$_{0.67}$Se$_{0.33}$. 
}
\label{fig:T1log}
\end{figure}

In Fig. \ref{fig:T1log}, we present the temperature dependence of the $^{125}$Te nuclear spin-lattice relaxation rate 
($1/T_{1}$) in Fe$_{1.04}$Se$_{0.33}$Te$_{0.67}$ for both conditions of $H // a$ and $H // c$. 
The recovery of the nuclear spin momentum cannot be explained with single $T_{1}$ component, and was   
fitted with the sum of two $T_{1}$ components of single-exponential. 
This is similar to the case with FeSe$_{0.92}$ \cite{Masaki} while nearly stoichiometric 
Fe$_{1.01}$Se does not show any distributions of $T_{1}$ \cite{Imai}. In the normal 
state, we estimated short and long components of $T_{1}$ by fitting the recoveries to two $T_{1}$ 
components of single-exponential for recovery curves measured at about $H \sim 8$ T. 
The main part of the recovery comes from the short $T_{1}$ component with the fraction of $\sim 80 \%$.
 In the superconducting state, we also measured $T_{1}$ at more lower field ($H 
\sim 4.3$) T. The recovery curves below $T\mathrm{_{c}}$ have large distribution 
of $T_{1}$ due to contributions of superconducting and vortex sites, and cannot be 
fitted with two components. In these regions, we estimated $T_{1}$ from the main 
component of the recovery curves, then the main component of $T_{1}$ below 8 K is 
slowest component which should come from the intrinsic superconductivity. 

In the normal state, the relaxation rates of both short and long $T_{1}$ components, 
$1/T_{1S}$ and $1/T_{1L}$, decrease with decreasing temperature, 
and the anisotropy is small. For the main short $T_{1}$ component, 
$1/T_{1S, H//a}$ is about 20 $\%$ larger than $1/T_{1S, H//c}$. 
$1/T_{1S, H//a}$ and $1/T_{1S, H//c}$ are proportional to $(\delta 
h_{a})^{2} + (\delta h_{c})^{2}$ and $(\delta h_{a})^{2} + (\delta h_{a})^{2}$, 
respectively, where $\delta h_{a, c}$ are local magnetic fluctuations. Taking it into account 
that $^{125}A_{//c}$ is larger than $^{125}A_{//a}$, the anisotropy of 
$1/T_{1}$ can be qualitatively explained by the anisotropy of the hyperfine 
coupling constant estimated from $K - \chi$ plots. 

 Below $T\mathrm{_{c}}$, $1/T_{1}$ decreases roughly proportional to $\sim 
T^{3}$ without coherent peak, resulting in an anisotropic superconductivity with line 
node. At low temperatures, the slope seems to close with $T$-linear due to the residual 
density of state, indicating the presence of node in the superconducting gap as well. In 
addition to the fact that the spin part of the Knight shift is suppressed in both $H // a$ and $H // c$, 
our $1/T_{1}$ results are consistent with the case of the $d$-wave superconductivity. We 
need more precise studies to elucidate the symmetry of the superconducting gap. 

%fig 5
\begin{figure}[tb]
\begin{center}
\includegraphics[width=7cm]{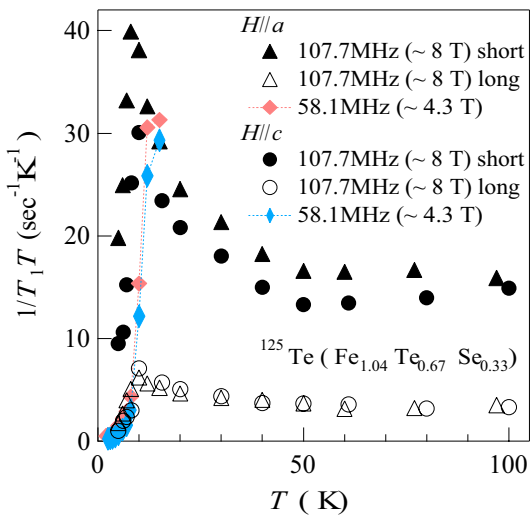}
\end{center}
\caption{
(a)Temperature dependence of the nuclear spin-lattice relaxation rate divided by temperature  
($1/T_{1}T$) for $^{125}$Te in Fe$_{1.04}$Te$_{0.67}$Se$_{0.33}$. 
}
\label{fig:invT1T}
\end{figure}

 Figure \ref{fig:invT1T} shows the temperature dependence of the nuclear spin-lattice relaxation rate 
divided by temperature  ($1/T_{1}T$) of $^{125}$Te in 
Fe$_{1.04}$Te$_{0.67}$Se$_{0.33}$. In both conditions of $H // a$ and $H // c$, both 
$1/T_{1S}$ of the main short $T_{1}$ component and 
$1/T_{1L}$ of the long $T_{1}$ component are enhanced with the decrease of 
temperature. $1/T_{1}T$ is proportional to $\Sigma_{\textbf{q}}|A_{\textrm{hf}}(\textbf{q})|^{2}\chi''(\textbf{q},\omega_n)/\omega_n$, where $A_{\textbf{q}}$,  $\chi''(\textbf{q},\omega_n)$ and $\omega_n$ are the $\textbf{q}$ depending hyperfine coupling constant, the imaginary part of the dynamical spin susceptibility at $\omega_n$, and NMR frequency, respectively\cite{Moriya_T1}.The enhancement of  $1/T_{1}T$ is attributed 
to the development of $\chi''(\textbf{q},\omega_n)$. The tendency of the 
enhancement of $1/T_{1}T$ at low temperatures is similar to the case with 
Fe$_{1.01}$Se \cite{Imai}

 In the earlier reported neutron diffraction, an incommensurate $(\delta\pi, 
\delta\pi)$ short-range magnetic ordering was observed even in the superconducting 
Fe$_{1.080}$Te$_{0.67}$Se$_{0.33}$, and $\delta$ can be tunable with the amount 
of excess irons \cite{Bao}. The spin fluctuations with incommensurate $\textbf{q}$ are 
expected to give a finite $A(\textbf{q})$  at any crystal sites. Therefore, the 
enhancement of $1/T_{1}T$ in our result may be probed as the similar spin fluctuations 
revealed by neutron diffraction. In our preliminary measurements, the absolute value of 
$1/T_{1}T$ in Fe$_{1.04}$Te$_{0.67}$Se$_{0.33}$ is larger than that in 
Fe$_{1.12}$Te$_{0.72}$Se$_{0.28}$ which has lager amount of excess irons and 
shows very small amount of superconducting volume fraction. Therefore, the increase 
of excess ions should decrease the density of states (DOS) at Fermi surface. By the 
density function calculations, the curious valence state Fe$^{+}$ is predicted to occur 
in excess irons sites, i.e.,  Fe(I\hspace{-0.1em}I) sites in Fe$_{1+\delta}$Te with nearly localized strong 
magnetic moment \cite{Zhang}. The role of excess irons is not only pair breaking 
magnetic scattering at the superconducting state but also causes reduction of DOS by 
electron doping, making the superconducting state unstable. Since the 
superconductivity strongly relates to the magnetic instability, the magnetic fluctuations 
proved by $1/T_{1}T$ are thought to be the driving force and origin of superconductivity 
in this Fe based system. 

We measured $^{125}$Te NMR on the single crystal of the novel layered 
superconductor Fe$_{1.04}$Se$_{0.33}$Te$_{0.67}$. We found that the 
superconducting gap has the line node with the spin singlet superconducting pairing. We also 
found that the magnetic fluctuations have an important role for the occurrence of the 
superconductivity. 

This work was supported by Grants-in-Aid for Scientific Research on Priority Area 
"Invention of anomalous quantum materials", from the Ministry of Education, Culture, 
Sports, Science and Technology of Japan (16076210) and also by Grants-in-Aid for 
Scientific Research from the Japan Society for Promotion of Science (19350030).

\end{document}